\documentclass[aps,preprint,nofootinbib,showpacs]{revtex4}%
\usepackage{amsfonts}
\usepackage{amsmath}
\usepackage{amssymb}
\usepackage{graphicx}
\usepackage[usenames,dvipsnames]{color}%
\providecommand{\U}[1]{\protect \rule{.1in}{.1in}}

\begin{document}
\preprint{ }
\title[Optical conductivity of a strong-coupling polaron]{Optical conductivity of a strong-coupling polaron}
\author{S. N. Klimin}
\altaffiliation{On leave from Department of Theoretical Physics, State University of Moldova,
str. A. Mateevici 60, MD-2009 Kishinev, Republic of Moldova.}

\author{J. T. Devreese}
\altaffiliation{Also at Technische Universiteit Eindhoven, P.O. Box 513, 5600 MB Eindhoven,
The Netherlands.}

\affiliation{Theorie van Kwantumsystemen en Complexe Systemen, Universiteit Antwerpen,
Universiteitsplein 1, B-2610 Antwerpen, Belgium}


\begin{abstract}
The polaron optical conductivity is derived within the strong-coupling
expansion, which is asymptotically exact in the strong-coupling limit. The
polaron optical conductivity band is provided by the multiphonon optical
transitions. The polaron optical conductivity spectra calculated within our
analytic strong-coupling approach and the numerically accurate Diagrammatic
Quantum Monte Carlo (DQMC) data are in a good agreement with each other at
large $\alpha \gtrapprox9$.

\end{abstract}
\volumeyear{2009}
\volumenumber{number}
\issuenumber{number}
\eid{identifier}
\date{\today}
\startpage{1}
\endpage{ }
\maketitle

\section{Introduction \label{sec:intro}}

The optical conductivity of the Fr\"{o}hlich polaron model attracted attention
for years \cite{Devreese2009}. In the regime of weak coupling, the optical
absorption of a polaron was calculated using different methods, e. g., Green's
function method \cite{GLF1962}, the Low-Lee-Pines formalism
\cite{DHL1971,HD1973}, perturbation expansion of the current-current
correlation function \cite{Sernelius1993}. The strong-coupling polaron optical
conductivity was calculated taking into account one-phonon \cite{KED1969} and
two-phonon \cite{GSD1973} transitions from the polaron ground state to the
polaron relaxed excited state (RES). In fact the present work finalizes the
project started in Ref. \cite{KED1969}. Using the path integral response
formalism, the impedance function of an all-coupling polaron was calculated by
FHIP \cite{FHIP1962} on the basis of the Feynman polaron model
\cite{Feynman1955}. Developing further the FHIP approach, the optical
conductivity was calculated in the path-integral formalism at zero temperature
\cite{DSG1972} and at finite temperatures \cite{PD1983}. In Ref.
\cite{DeFilippis2006}, the extension of the method of Ref. \cite{DSG1972}
accounting for the polaron damping (for the polaron coupling constant
$\alpha \lesssim8$) and the asymptotic strong-coupling approach using the
Franck-Condon (FC) picture for the optical conductivity (for $\alpha \gtrsim8$)
have given reasonable results for the polaron optical conductivity at all
values of $\alpha$. The concept of the RES and FC polaron states played a key
role in the understanding of the mechanism of the polaron optical conductivity
\cite{KED1969,GSD1973,DE1964,DSG1972,Devreese1972,PD1983}.

Recently, the Diagrammatic Quantum Monte Carlo (DQMC) numerical method has
been developed \cite{Mishchenko2000,Mishchenko2003}, which provides accurate
results for the polaron characteristics in all coupling regimes. The analytic
treatment \cite{DSG1972} was intended to be valid at all coupling strengths.
However, it is established in \cite{DSG1972,Devreese1972,GSD1973} that the
linewidth of the obtained spectra \cite{DSG1972} is unreliable for
$\alpha \gtrapprox7$. Nevertheless, the position of the peak attributed to RES
in Ref. \cite{DSG1972} is close to the maximum of the polaron optical
conductivity band calculated using DQMC up to very large values of $\alpha$
(see Fig. 1).

An extension of the path-integral approach \cite{DSG1972} performed in Ref.
\cite{DeFilippis2006} gives a good agreement with DQMC for weak and
intermediate coupling strengths. In the strong-coupling limit, in Ref.
\cite{DeFilippis2006} the adiabatic strong-coupling expansion was applied.
That expansion, however, is not exact in the strong-coupling limit because of
a parabolic approximation \cite{LP1948} for the adiabatic potential.

In the present work, the strong-coupling approach of Ref.
\cite{DeFilippis2006} is extended in order to obtain the polaron optical
conductivity which is \emph{asymptotically exact in the strong-coupling
limit}. We develop the multiphonon strong-coupling expansion using numerically
accurate in the strong-coupling limit polaron energies and wave functions and
accounting for non-adiabaticity.

\section{Optical conductivity \label{sec:optcon}}

We consider the electron-phonon system with the Hamiltonian written down in
the Feynman units ($\hbar=1,$ the carrier band mass $m_{b}=1$, and the
LO-phonon frequency $\omega_{\mathrm{LO}}=1$)%
\begin{equation}
H=\frac{\mathbf{p}^{2}}{2}+\sum_{\mathbf{q}}\left(  b_{\mathbf{q}}%
^{+}b_{\mathbf{q}}+\frac{1}{2}\right)  +\frac{1}{\sqrt{V}}\sum_{\mathbf{q}%
}\frac{\sqrt{2\sqrt{2}\pi \alpha}}{q}\left(  b_{\mathbf{q}}+b_{-\mathbf{q}}%
^{+}\right)  e^{i\mathbf{q\cdot r}}. \label{H}%
\end{equation}
where $\mathbf{r},\mathbf{p}$ represent the position and momentum of an
electron, $b_{\mathbf{q}}^{+},b_{\mathbf{q}}$ denote the creation and
annihilation operators for longitudinal optical (LO) phonons with wave vector
$\mathbf{q}$, and $V_{\mathbf{q}}$ describes the amplitude of the interaction
between the electrons and the phonons. For the Fr\"{o}hlich electron-phonon
interaction, the amplitude of the electron -- LO-phonon interaction is%
\begin{equation}
V_{\mathbf{q}}=\frac{1}{\sqrt{V}}\frac{\sqrt{2\sqrt{2}\pi \alpha}}{q}
\label{Vq}%
\end{equation}
with the crystal volume $V$, and the electron-phonon coupling constant
$\alpha$.

The polaron optical conductivity describes the response of the system with the
Hamiltonian (\ref{H}) to an applied electromagnetic field (along the $z$-axis)
with frequency $\omega$. This optical response is expressed using the Kubo
formula with a dipole-dipole correlation function:
\begin{equation}
\operatorname{Re}\sigma \left(  \omega \right)  =\frac{n_{0}\omega}{2}\left(
1-e^{-\beta \omega}\right)  \int_{-\infty}^{\infty}e^{i\omega t}\left \langle
d_{z}\left(  t\right)  d_{z}\right \rangle \,dt, \label{KuboDD}%
\end{equation}
where $\mathbf{d}=-e_{0}\mathbf{r}$ is the electric dipole moment, $e_{0}$ is
the unit charge, $\beta=\frac{1}{k_{B}T}$, $n_{0}$ is the electron density. In
the zero-temperature limit, the optical conductivity (\ref{KuboDD}) measured
in units of $e_{0}^{2}$ becomes%
\begin{equation}
\operatorname{Re}\sigma \left(  \omega \right)  =\frac{\omega}{2}\int_{-\infty
}^{\infty}e^{i\omega t}f_{zz}\left(  t\right)  \,dt, \label{KuboDD0}%
\end{equation}
with the correlation function
\begin{equation}
f_{zz}\left(  t\right)  \equiv \left \langle z\left(  t\right)  z\left(
0\right)  \right \rangle =\left \langle \Psi_{0}\left \vert e^{itH}%
ze^{-itH}z\right \vert \Psi_{0}\right \rangle , \label{fzz}%
\end{equation}
where $\left \vert \Psi_{0}\right \rangle $ is the ground-state wave function of
the electron-phonon system.

Within the strong-coupling approach, the ground-state wave function is chosen
as the product of a trial wave function of an electron $\left \vert \psi
_{0}^{\left(  e\right)  }\right \rangle $ and of a trial wave function of a
phonon subsystem $\left \vert \Phi_{ph}\right \rangle $:%
\begin{equation}
\left \vert \Psi_{0}\right \rangle =\left \vert \psi_{0}^{\left(  e\right)
}\right \rangle \left \vert \Phi_{ph}\right \rangle . \label{ansatz}%
\end{equation}
The phonon trial wave function is written as the strong-coupling unitary
transformation applied to the phonon vacuum
\begin{equation}
\left \vert \Phi_{ph}\right \rangle =U\left \vert 0_{ph}\right \rangle .
\label{Fph}%
\end{equation}
with the unitary operator%
\begin{equation}
U=e^{\sum_{\mathbf{q}}\left(  f_{\mathbf{q}}b_{\mathbf{q}}-f_{\mathbf{q}%
}^{\ast}b_{\mathbf{q}}^{+}\right)  }, \label{unit}%
\end{equation}
and the variational parameters $\left \{  f_{\mathbf{q}}\right \}  $. The
transformed Hamiltonian $\tilde{H}\equiv U^{-1}HU$ takes the form%
\begin{equation}
\tilde{H}=\tilde{H}_{0}+W \label{HT2}%
\end{equation}
with the terms%
\begin{align}
\tilde{H}_{0}  &  =\frac{\mathbf{p}^{2}}{2}+\sum_{\mathbf{q}}\left \vert
f_{\mathbf{q}}\right \vert ^{2}+V_{a}\left(  r\right)  +\sum_{\mathbf{q}%
}\left(  b_{\mathbf{q}}^{+}b_{\mathbf{q}}+\frac{1}{2}\right)  ,\label{H0}\\
W  &  =\sum_{\mathbf{q}}\left(  W_{\mathbf{q}}b_{\mathbf{q}}+W_{\mathbf{q}%
}^{\ast}b_{\mathbf{q}}^{+}\right)  . \label{W}%
\end{align}
Here, $W_{\mathbf{q}}$ are the amplitudes of the renormalized electron-phonon
interaction%
\begin{equation}
W_{\mathbf{q}}=\frac{\sqrt{2\sqrt{2}\pi \alpha}}{q\sqrt{V}}\left(
e^{i\mathbf{q\cdot r}}-\rho_{\mathbf{q}}\right)  ,
\end{equation}
where $\rho_{\mathbf{q}}$ is the expectation value of the operator
$e^{i\mathbf{q\cdot r}}$ with the trial electron wave function $\left \vert
\psi_{0}^{\left(  e\right)  }\right \rangle $:%
\begin{equation}
\rho_{\mathbf{q}}=\left \langle \psi_{0}^{\left(  e\right)  }\left \vert
e^{i\mathbf{q\cdot r}}\right \vert \psi_{0}^{\left(  e\right)  }\right \rangle ,
\end{equation}
and $V_{a}\left(  r\right)  $ is the self-consistent potential energy for the
electron,%
\begin{equation}
V_{a}\left(  r\right)  =-\sum_{\mathbf{q}}\frac{4\sqrt{2}\pi \alpha}{q^{2}%
V}\rho_{-\mathbf{q}}e^{i\mathbf{q}\cdot \mathbf{r}}.
\end{equation}

Averaging the Hamiltonian (\ref{HT2}) with the phonon vacuum $\left \vert
0\right \rangle $ and with the trial electron wave function $\left \vert
\psi_{0}\right \rangle $, we arrive at the following variational expression for
the ground-state energy%
\begin{align}
E_{0}  &  =\left \langle \Psi_{0}\left \vert H\right \vert \Psi_{0}\right \rangle
=\left \langle \psi_{0}\left \vert \frac{\mathbf{p}^{2}}{2}\right \vert \psi
_{0}\right \rangle +\sum_{\mathbf{q}}\left \vert f_{\mathbf{q}}\right \vert
^{2}\nonumber \\
&  -\sum_{\mathbf{q}}\left(  V_{\mathbf{q}}f_{\mathbf{q}}^{\ast}%
\rho_{\mathbf{q}}+V_{\mathbf{q}}^{\ast}f_{\mathbf{q}}\rho_{-\mathbf{q}%
}\right)  , \label{E0var}%
\end{align}
After minimization of the polaron ground-state energy (\ref{E0var}), the
parameters $f_{\mathbf{q}}$ acquire their optimal values%
\begin{equation}
f_{\mathbf{q}}=V_{\mathbf{q}}\rho_{\mathbf{q}}. \label{ov}%
\end{equation}
The ground-state energy with $\left \{  f_{\mathbf{q}}\right \}  $ given by Eq.
(\ref{ov}) takes the form%
\begin{equation}
E_{0}=\left \langle \psi_{0}\left \vert \frac{\mathbf{p}^{2}}{2}\right \vert
\psi_{0}\right \rangle -\sum_{\mathbf{q}}\left \vert V_{\mathbf{q}}\right \vert
^{2}\left \vert \rho_{\mathbf{q}}\right \vert ^{2}. \label{E0}%
\end{equation}

With the strong-coupling Ansatz (\ref{ansatz}) for the polaron ground-state
wave function and after the application of the unitary transformation
(\ref{unit}), the correlation function (\ref{fzz}) takes the form%
\begin{equation}
f_{zz}\left(  t\right)  =\left \langle 0_{ph}\left \vert \left \langle \psi
_{0}\left \vert e^{it\tilde{H}}ze^{-it\tilde{H}}z\right \vert \psi
_{0}\right \rangle \right \vert 0_{ph}\right \rangle . \label{fzz2}%
\end{equation}

This correlation function can be expanded using a complete orthogonal set of
intermediate states $\left \vert j\right \rangle $ and the completeness
property:%
\begin{equation}
\sum_{j}\left \vert j\right \rangle \left \langle j\right \vert =1. \label{compl}%
\end{equation}
In the present work, we use the intermediate basis of the Franck-Condon (FC)
states. The FC states correspond to the equilibrium phonon configuration for
the ground state. Thus the FC wave functions are the exact eigenstates of the
Hamiltonian $\tilde{H}_{0}$. Further on, the FC wave functions are written in
the spherical-wave representation as $\left \vert \psi_{n,l,m}\right \rangle
=R_{n,l}\left(  r\right)  Y_{l,m}\left(  \theta,\varphi \right)  $ where
$R_{n,l}\left(  r\right)  $ are the radial wave functions, and $Y_{l,m}\left(
\theta,\varphi \right)  $ are the spherical harmonics, $l$ is the quantum
number of the angular momentum, $m$ is the $z$-projection of the angular
momentum, and $n$ is the radial quantum number\footnote{In this
classification, the ground-state wave function is $\left \vert \psi
_{0,0,0}\right \rangle \equiv \left \vert \psi_{0}\right \rangle $.}. The energy
levels for the eigenstates of the Hamiltonian $\tilde{H}_{0}$ are denoted
$E_{n,l}$.

Using (\ref{compl}) with that complete and orthogonal basis , we transform
(\ref{fzz2}) to the expression%
\begin{align}
f_{zz}\left(  t\right)   &  =\sum_{\substack{n,l,m,\\n^{\prime},l^{\prime
},m^{\prime},\\n^{\prime \prime},l^{\prime \prime},m^{\prime \prime}%
}}\left \langle \psi_{n,l,m}\left \vert z\right \vert \psi_{n^{\prime \prime
},l^{\prime \prime},m^{\prime \prime}}\right \rangle \left \langle \psi
_{n^{\prime},l^{\prime},m^{\prime}}\left \vert z\right \vert \psi_{0}%
\right \rangle \nonumber \\
&  \times \left \langle 0_{ph}\left \vert \left \langle \psi_{0}\left \vert
e^{it\tilde{H}}\right \vert \psi_{n,l,m}\right \rangle \left \langle
\psi_{n^{\prime \prime},l^{\prime \prime},m^{\prime \prime}}\left \vert
e^{-it\tilde{H}}\right \vert \psi_{n^{\prime},l^{\prime},m^{\prime}%
}\right \rangle \right \vert 0_{ph}\right \rangle . \label{fzz3}%
\end{align}

So far, the only approximation made in (\ref{fzz3}) is the strong-coupling
Ansatz for the polaron ground-state wave function. However, in order to obtain
a numerically tractable expression for the polaron optical conductivity, an
additional approximation valid in the strong-coupling limit must be applied to
the matrix elements of the evolution operator $e^{-it\tilde{H}}$ with the
Hamiltonian of the electron-phonon system $\tilde{H}$ given by formula
(\ref{HT2}). According to Ref. \cite{Allcock1}, in the strong-coupling limit,
the matrix elements of the Hamiltonian of the electron-phonon system between
states corresponding to different energy levels are of order of magnitude
$\alpha^{-4}$. Therefore in the strong-coupling regime these matrix elements
can be neglected; this is called the adiabatic or the Born-Oppenheimer (BO)
approximation \cite{Allcock1}, because of its strict analogy with the
Born-Oppenheimer adiabatic approximation in the theory of molecules and
crystals (\cite{Born}, p. 171). Consequently, in the further treatment we
neglect the matrix elements $\left \langle \psi_{n,l,m}\left \vert
e^{-it\tilde{H}}\right \vert \psi_{n^{\prime},l^{\prime},m^{\prime}%
}\right \rangle $ for the FC states with different energies, $E_{n,l}\neq
E_{n^{\prime},l^{\prime}}$. The same scheme was used in the theory of the
multi-phonon optical processes for bound electrons interacting with phonons
\cite{Perlin,Pekar1954}.

Strictly speaking, the summation over the excited polaron states in Eq.
(\ref{fzz3}) must involve the transitions to both the discrete and continuous
parts of the polaron spectrum. A transition to the states of the continuous
spectrum means that the electron leaves the polaron potential well. Therefore
these transitions can be attributed to the \textquotedblleft polaron
dissociation\textquotedblright. The transitions to the continuous spectrum are
definitely beyond the adiabatic approximation. As shown in Ref.
\cite{Pekar1954}, the transition probability to the states of the continuous
spectrum is very small compared with the transition probability between the
ground and the first excited state (which belongs to the discrete part of the
polaron energy spectrum). We neglect here the contribution to the polaron
optical conductivity due to the transitions to the continuous spectrum.

The matrix elements neglected within the adiabatic approximation correspond to
the transitions between FC states with different energies due to the
electron-phonon interaction. Hence these transitions can be called
non-adiabatic. The adiabatic approximation is related to the matrix elements
of the evolution operator $e^{-it\tilde{H}}$. On the contrary, the matrix
elements of the transitions between different FC states for the electric
dipole moment are, in general, not equal to zero. Moreover, these transitions
can be accompanied by the emission of phonons. The electron FC wave functions
constitute a complete orthogonal set. However, the corresponding phonon wave
functions can be non-orthogonal because of a different shift of phonon
coordinates for different electron states. This makes multi-phonon transitions
possible \cite{Perlin}. It is important to note that in our treatment we
neglect only the non-adiabatic transitions between the electron states with
\emph{different} energies. On the contrary, the transitions within one and the
same degenerate level can be non-adiabatic. This \emph{internal
non-adiabaticity} (i.~e., the non-adiabaticity of the transitions within one
and the same degenerate level) is taken into account in the subsequent treatment.

It is useful to stress the difference between the strong-coupling Ansatz and
the adiabatic approximation. The strong-coupling Ansatz consists of the choice
of the trial variational ground state wave function for the electron-phonon
system in the factorized form (\ref{ansatz}). The adiabatic approximation
means neglecting the matrix elements of the evolution operator between
internal polaron states with different energies. These two approximations are
not the same, but they both are valid in the strong-coupling regime and
consistent with each other.

The correlation function (\ref{fzz3}) is transformed in the following way. The
exponents $e^{it\tilde{H}}$ and $e^{-it\tilde{H}}$ are disentangled:%
\begin{align}
e^{-it\tilde{H}}  &  =e^{-it\tilde{H}_{0}}\mathrm{T}\exp \left(  -i\int_{0}%
^{t}dsW\left(  s\right)  \right)  ,\\
e^{it\tilde{H}}  &  =e^{it\tilde{H}_{0}}\mathrm{T}\exp \left(  i\int_{0}%
^{t}dsW\left(  -s\right)  \right)
\end{align}
where $W\left(  s\right)  $ is the renormalized electron-phonon interaction
Hamiltonian $W$ in the interaction representation,%
\begin{equation}
W\left(  s\right)  \equiv e^{is\tilde{H}_{0}}We^{-is\tilde{H}_{0}}.
\end{equation}
This gives us the result%
\begin{align}
&  f_{zz}\left(  t\right)  =\sum_{\substack{n,l,m,\\n^{\prime},l^{\prime
},m^{\prime},\\n^{\prime \prime},l^{\prime \prime},m^{\prime \prime}%
}}\left \langle \psi_{n,l,m}\left \vert z\right \vert \psi_{n^{\prime \prime
},l^{\prime \prime},m^{\prime \prime}}\right \rangle \left \langle \psi
_{n^{\prime},l^{\prime},m^{\prime}}\left \vert z\right \vert \psi_{0}%
\right \rangle e^{it\left(  E_{0}-E_{n^{\prime \prime},l^{\prime \prime}}\right)
}\nonumber \\
&  \times \left \langle 0_{ph}\left \vert \left \langle \psi_{0}\left \vert
\mathrm{T}\exp \left(  i\int_{0}^{t}dsW\left(  -s\right)  \right)  \right \vert
\psi_{n,l,m}\right \rangle \right.  \right. \nonumber \\
&  \left.  \left.  \times \left \langle \psi_{n^{\prime \prime},l^{\prime \prime
},m^{\prime \prime}}\left \vert \mathrm{T}\exp \left(  -i\int_{0}^{t}dsW\left(
s\right)  \right)  \right \vert \psi_{n^{\prime},l^{\prime},m^{\prime}%
}\right \rangle \right \vert 0_{ph}\right \rangle . \label{fzz4}%
\end{align}

Within the adiabatic approximation, the optical conductivity is simplified.
The full details of the derivation are described in the Appendix A. First,
using the selection rules for the dipole matrix elements, the spherical
symmetry of the Hamiltonian $\tilde{H}$ and the adiabatic approximation, the
correlation function (\ref{fzz4}) is reduced to the form%
\begin{align}
f_{zz}\left(  t\right)   &  =\sum_{n}D_{n}e^{-i\Omega_{n,0}t}\nonumber \\
&  \times \left \langle \psi_{n,1,0}\left \vert \left \langle 0_{ph}\left \vert
\mathrm{T}\exp \left[  -i\int_{0}^{t}dsW\left(  s\right)  \right]  \right \vert
0_{ph}\right \rangle \right \vert \psi_{n,1,0}\right \rangle \label{fzz4a}%
\end{align}
where $\Omega_{n,0}$ is the FC transition frequency%
\begin{equation}
\Omega_{n,0}\equiv E_{n,1}-E_{0}, \label{Wn0}%
\end{equation}
and $D_{n}$ is the squared modulus of the dipole transition matrix element%
\begin{equation}
D_{n}=\left \vert \left \langle \psi_{0}\left \vert z\right \vert \psi
_{n,1,0}\right \rangle \right \vert ^{2}. \label{Dn}%
\end{equation}

Within the adiabatic approximation, the partial (with the electron wave
functions) averaging of the operator T-exponent in (\ref{fzz4a}) can be
exactly performed (see details in Appendix A). As a result, the optical
conductivity is transformed to the expression%
\begin{align}
\operatorname{Re}\sigma \left(  \omega \right)   &  =\frac{\omega}{6}\sum
_{n}D_{n}\int_{-\infty}^{\infty}e^{i\left(  \omega-\Omega_{n,0}\right)
t}\nonumber \\
&  \times \left \langle 0_{ph}\left \vert \mathrm{Tr}\left(  \mathrm{T}%
\exp \left[  -i\int_{0}^{t}ds\mathbb{W}^{\left(  n\right)  }\left(  s\right)
\right]  \right)  \right \vert 0_{ph}\right \rangle dt. \label{f-ad3}%
\end{align}
The T-exponent in (\ref{f-ad3}) contains the finite-dimensional matrix
$\mathbb{W}^{\left(  n\right)  }\left(  s\right)  $ depending on the phonon
coordinates:%
\begin{equation}
\left(  \mathbb{W}_{k,l,m}^{\left(  n\right)  }\right)  _{m_{1},m_{2}%
}=\left \langle \psi_{n,1,m_{1}}\left \vert W_{k,l,m}\right \vert \psi
_{n,1,m_{2}}\right \rangle \label{W1}%
\end{equation}
where $W_{k,l,m}$ are the amplitudes of the electron-phonon interaction in the
basis of spherical wave functions.

Because the kinetic energy of the phonons is of order $\alpha^{-4}$ compared
to the leading term of the Hamiltonian \cite{Allcock1}, we neglect this
kinetic energy in the present work, because the treatment is related to the
strong-coupling regime. As a result, $Q_{k,l,m}$ commute with the Hamiltonian
$\tilde{H}_{0},$ so that in (\ref{f-ad3}), $\mathbb{W}^{\left(  n\right)
}\left(  s\right)  =\mathbb{W}^{\left(  n\right)  }$. Furthermore, in a
finite-dimensional basis $\left \{  \left \vert \psi_{n,l,m}\right \rangle
\right \}  $ for a given level $\left(  n,l\right)  $, all eigenvalues of the
Hamiltonian $\tilde{H}_{0}$ are the same. Therefore the $\mathrm{T}$-exponent
entering (\ref{f-ad3}) in that finite-dimensional basis turns into a usual
exponent. As a result, the strong-coupling polaron optical conductivity
(\ref{f-ad3}) takes the form%
\begin{equation}
\operatorname{Re}\sigma \left(  \omega \right)  =\frac{\omega}{6}\sum_{n}%
D_{n}\int_{-\infty}^{\infty}e^{i\left(  \omega-\Omega_{n,0}\right)
t}\left \langle 0_{ph}\left \vert \mathrm{Tr}\exp \left(  -i\mathbb{W}^{\left(
n\right)  }t\right)  \right \vert 0_{ph}\right \rangle dt. \label{fzz1}%
\end{equation}

The matrix interaction Hamiltonian (\ref{W1}) depends on the phonon
coordinates, and the matrices $\mathbb{W}_{k,l,m}^{\left(  n\right)  }$ with
different $m$ for one and the same degenerate energy level do not commute with
each other. According to the Jahn -- Teller theorem \cite{JT}, for a
degenerate level there does not exist a unitary transformation which
simultaneously diagonalizes all matrices $\mathbb{W}_{k,l,m}^{\left(
n\right)  }$ in a basis that does not depend on the phonon coordinates. The
manifestations of that theorem are attributed to the Jahn -- Teller effect.
Therefore, because we neglect the non-commutation of the matrices
$\mathbb{W}_{k,l,m}^{\left(  n\right)  }$, the Jahn -- Teller effect is omitted.

In fact, neglecting the Jahn -- Teller effect is not necessary. The averaging
in Eq. (\ref{fzz1}) is performed exactly using the effective phonon modes
similarly to Ref. \cite{Lumin1998} (see the details in Appendix B). As a
result, we arrive at the following expression for the strong-coupling polaron
optical conductivity%
\begin{align}
\operatorname{Re}\sigma \left(  \omega \right)   &  =\frac{\omega}{3\pi^{2}}%
\sum_{n}\frac{D_{n}}{a_{0}^{\left(  n\right)  }}\int_{-\infty}^{\infty}%
dx_{0}\int_{-\infty}^{\infty}dx_{1}\int_{-\infty}^{\infty}dx_{2}\int_{-\infty
}^{\infty}dy_{1}\int_{-\infty}^{\infty}dy_{2}\nonumber \\
&  \times \sum_{j=1}^{3}\exp \left \{  -\frac{1}{2}\left[  x_{0}^{2}+\sum
_{m=1,2}\left(  x_{m}^{2}+y_{m}^{2}\right)  +\frac{\left(  \omega-\Omega
_{n,0}-\frac{a_{2}^{\left(  n\right)  }}{2\sqrt{5\pi}}\lambda_{j}\left(
Q_{2}\right)  \right)  ^{2}}{\left(  a_{0}^{\left(  n\right)  }\right)  ^{2}%
}\right]  \right \}  . \label{Resig}%
\end{align}
Here, $\lambda_{j}\left(  Q_{2}\right)  $ are the eigenvalues for the matrix
interaction Hamiltonian, which are explicitly determined in the Appendix B by
the formula (\ref{eigenv}). The coefficients $a_{0}^{\left(  n\right)  }$ and
$a_{2}^{\left(  n\right)  }$ are given by (\ref{c0n}) and (\ref{c2n}),
respectively. The polaron optical conductivity given by the expression
(\ref{Resig}), is in fact an envelope of the multiphonon polaron optical
conductivity band with the correlation function (\ref{f-ad3}) provided by the
phonon-assisted transitions from the polaron ground state to the polaron RES.
This result is consistent with Ref. \cite{KED1969}, where the same paradigm of
the phonon-assisted transitions to the polaron RES was exploited, but the
calculation was limited to the one-phonon transition.

In order to reveal the significance of the Jahn -- Teller effect for the
polaron, we alternatively calculate $\left \langle 0_{ph}\left \vert
\mathrm{Tr}\exp \left(  -i\mathbb{W}^{\left(  n\right)  }t\right)  \right \vert
0_{ph}\right \rangle $ neglecting the non-commutation of the matrices
$\mathbb{W}_{k,l,m}^{\left(  n\right)  }$, as described in the Appendix B.~2.
The resulting expression for the polaron optical conductivity is much simpler
than formula (\ref{Resig}) and is similar to the expression (3) of Ref.
\cite{DeFilippis2006}:%
\begin{equation}
\operatorname{Re}\sigma \left(  \omega \right)  =\omega \sum_{n}\sqrt{\frac{\pi
}{2\omega_{s}^{\left(  n\right)  }}}D_{n}\exp \left(  -\frac{\left(
\omega-\Omega_{n,0}\right)  ^{2}}{2\omega_{s}^{\left(  n\right)  }}\right)  ,
\label{SJT}%
\end{equation}
with the parameter (often called the Huang-Rhys factor)%
\begin{equation}
\omega_{s}^{\left(  n\right)  }=\frac{1}{2}\left(  a_{0}^{\left(  n\right)
}\right)  ^{2}+\frac{1}{4\pi}\left(  a_{2}^{\left(  n\right)  }\right)  ^{2}.
\label{S}%
\end{equation}

The strong-coupling electron energies and wave functions in Eq. (\ref{f-ad3})
can be calculated using different approximations. For example, within the
Landau-Pekar (LP) approximation \cite{LP1948}, the trial wave function
$\left \vert \psi_{0}\right \rangle $ is chosen as the ground state of a 3D
oscillator. Within the Pekar approximation \cite{Pekar1954}, $\left \vert
\psi_{0}\right \rangle $ is chosen in the form
\begin{equation}
\left \vert \psi_{0}\left(  r\right)  \right \rangle =Ce^{-ar}\left(
1+ar+br^{2}\right)  \label{P}%
\end{equation}
with the variational parameters $a$ and $b$. Finally, the trial ground state
wave function can be determined numerically exactly following Miyake
\cite{Miyake1975} (see also \cite{Kleinert}, Chap. 5.22). Within the LP
approximation, formula (\ref{SJT}) reproduces the polaron optical conductivity
obtained in Ref. \cite{DeFilippis2006}.

In the LP approximation, the matrix elements $\left \langle \psi_{0}\left \vert
z\right \vert \psi_{n,1,0}\right \rangle $ are different from zero only for
$n=1$, i. e. only for the $1s\rightarrow2p$ transition. Beyond the LP
approximation, also the transitions to other excited states are allowed
because of the nonparabolicity of the self-consistent potential $V_{a}\left(
r\right)  $. The use of exact strong-coupling wave functions, instead of the
LP wave functions, may significantly influence the optical conductivity. In
the present treatment we use the numerically exact electron energies and wave
functions of both ground and first excited states according to Ref.
\cite{Miyake1975}. The FC transition energies $\Omega_{n,0}$ to leading order
of the strong-coupling approximation are determined according to (\ref{Wn0}).
In order to account for the corrections of the FC energy with accuracy up to
$\alpha^{0}$, we add to $\Omega_{n,0}$ the correction $\Delta \Omega
_{\mathrm{FC}}\approx-3.8$ from Ref. \cite{DeFilippis2006}. Because we use the
numerically accurate strong-coupling wave functions and energies corresponding
to Miyake \cite{Miyake1975}, the formula (\ref{fzz4}) \emph{is asymptotically
exact in the strong-coupling limit, at least in its leading term in powers of}
$\alpha^{-2}$.

\section{Results and discussion \label{sec:results}}

In Figs. 2 to 3, we have plotted the polaron optical conductivity spectra
calculated for different values of the coupling constant $\alpha$. The optical
conductivity spectra calculated within the present strong-coupling approach
taking into account the Jahn -- Teller effect are shown by the solid curves.
The optical conductivity derived neglecting the Jahn -- Teller effect is shown
by the dashed curves. It is worth mentioning that there is little difference
in the optical conductivity spectra between those calculated with and without
the Jahn -- Teller effect. The optical conductivity obtained in Ref.
\cite{DeFilippis2006} with the Landau-Pekar (LP) adiabatic approximation is
plotted with dash-dotted curves. The full dots show the numerical Diagrammatic
Quantum Monte Carlo (DQMC) data \cite{Mishchenko2003,DeFilippis2006}. The FC
transition frequency for the transition to the first excited FC state
$\Omega_{1,0}\equiv \Omega_{\mathrm{FC}}$ and the RES transition frequency
$\Omega_{\mathrm{RES}}$ are explicitly indicated in the figures.

The polaron optical conductivity spectra calculated within the present
strong-coupling approach are shifted to lower frequencies with respect to the
optical conductivity spectra calculated within the LP approximation of Ref.
\cite{DeFilippis2006}. This shift is due to the use of the numerically
accurate strong coupling energy levels and wave functions of the internal
polaron states, and of the numerically accurate self-consistent adiabatic
polaron potential.

According to the selection rules for the matrix elements of the
electron-phonon interaction, there is a contribution to the polaron optical
conductivity from the phonon modes with angular momentum $l=0$ ($s$-phonons)
and with angular momentum $l=2$ ($d$-phonons). The $s$-phonons are fully
symmetric, therefore they do not contribute to the Jahn -- Teller effect,
while the $d$-phonons are active in the Jahn -- Teller effect. The
contribution of the $d$-phonons to the optical conductivity spectra is not
small compared to the contribution of the $s$-phonons. However, the
distinction between the optical conductivity spectra calculated with and
without the Jahn -- Teller effect is relatively small.

For $\alpha=8$ and $\alpha=8.5$, the maxima of the polaron optical
conductivity spectra, calculated within the present strong-coupling approach
are positioned to the low frequency side of the maxima of those calculated
using the DQMC method. The agreement between our strong-coupling polaron
optical conductivity spectra and the numerical DQMC data improves with
increasing alpha. This is in accordance with the fact that the present
strong-coupling approach for the polaron optical conductivity is
asymptotically exact in the strong-coupling limit.

The total polaron optical conductivity must satisfy the sum rule
\cite{DLR1977}%
\begin{equation}
\int_{0}^{\infty}\operatorname{Re}\sigma \left(  \omega \right)  d\omega
=\frac{\pi}{2}. \label{sr1}%
\end{equation}
In the weak- and intermediate-coupling regimes at $T=0$, there are two
contributions to the left-hand side of that sum rule: (1) the contribution
from the polaron optical conductivity for $\omega>\omega_{\mathrm{LO}}$ and
(2) the contribution from the \textquotedblleft central peak\textquotedblright%
\ at $\omega=0$, which is proportional to the inverse polaron mass
\cite{DLR1977}. In the asymptotic strong-coupling regime, the inverse to the
polaron mass is of order $\alpha^{-4}$, and hence the contribution from the
\textquotedblleft central peak\textquotedblright \ to the polaron optical
conductivity is beyond the accuracy of the present approximation (where we
keep the terms $\propto \alpha^{-2}$ and $\propto \alpha^{0}$).

As discussed above, in the present work the transitions from the ground state
to the states of the continuous part of the polaron energy spectrum are
neglected. Therefore the integral over the frequency [the left-hand side of
(\ref{sr1})] for the optical conductivity calculated within the present
strong-coupling approximation can be (relatively slightly) smaller than
$\pi/2$. The relative contribution of the transitions to the continuous part
of the polaron spectrum, $\Delta_{c}$, can be therefore estimated as%
\begin{equation}
\Delta_{c}\equiv1-\frac{2}{\pi}\int_{0}^{\infty}\operatorname{Re}\sigma \left(
\omega \right)  d\omega, \label{Delta}%
\end{equation}
where the right-hand side is obtained by a numerical integration of
$\operatorname{Re}\sigma \left(  \omega \right)  $ calculated within the present
strong-coupling approach. This numeric estimation shows that for $\alpha>8$,
$\Delta_{c}<0.01$. Moreover, with increasing $\alpha$, the relative
contribution of the transitions to the continuous part of the polaron spectrum
falls down. This confirms the accuracy of the present strong-coupling approach.

In Refs. \cite{Emin1993,Myasnikov2006}, the optical conductivity of a
strong-coupling polaron was calculated assuming that in the strong-coupling
regime the polaron optical response is provided mainly by the transitions to
the continuous part of the spectrum (these transitions are called there
\textquotedblleft the polaron dissociation\textquotedblright). This concept is
in contradiction both with the early estimation by Pekar \cite{Pekar1954}
discussed above and with the very small weight of those transitions shown in
Fig. 4. The approach of Ref. \cite{Emin1993} in fact takes into account only a
small part of the strong-coupling polaron optical conductivity -- namely, the
high-frequency \textquotedblleft tail\textquotedblright \ of the optical
conductivity spectrum.

When comparing the polaron optical conductivity spectra calculated in the
present work with the DQMC data \cite{Mishchenko2003,DeFilippis2006}, we can
see that the present approach, with respect to DQMC, underestimates the
high-frequency part of the polaron optical conductivity. This difference,
however, gradually diminishes with increasing $\alpha$, in accordance with the
fact that the present method is an asymptotic strong-coupling approximation.

Because the optical conductivity spectra calculated in the present
strong-coupling approximation using the expressions (\ref{Resig}) and
(\ref{SJT}) represent the envelopes of the RES peak with the multi-phonon
satellites, the separate peeks are not explicitly seen in those spectra. The
FC and RES peaks are indicated in the figures by the arrows. The FC transition
frequency $\Omega_{1,0}$ in the strong-coupling case is positioned close to
the maximum of the polaron optical conductivity band (both calculated within
the present approach and within DQMC). The RES transition frequency is
positioned one $\omega_{\mathrm{LO}}$ below the onset of the LO-sidebands.
Note that the strong-coupling polaron optical conductivity derived in Refs.
\cite{Spohn1987} contains only the zero-phonon (RES) line and no phonon
satellites at all. In contrast, in the present calculation, the maximum of the
polaron optical conductivity spectrum shifts to higher frequencies with
increasing $\alpha$, so that the multiphonon processes invoking large number
of phonons become more and more important, in accordance with predictions of
Refs. \cite{KED1969,DSG1972}.

It is worth noting the following important point: the maximum of the polaron
optical conductivity band can be hardly interpreted as a broadened transition
to an FC state on the following reasons.\ Formula (\ref{f-ad3}) describes a
set of multi-phonon peaks. In the simplifying approximation which neglects the
Jahn -- Teller effect (see Ref. \cite{DeFilippis2006}), those peaks are
positioned at the frequencies $\omega=\tilde{\Omega}_{n,0}+k$, where $k$ is
the number of emitted phonons and is the frequency of the zero-phonon line.
The frequencies $\tilde{\Omega}_{n,0}$ do not coincide with the FC transition
frequencies but are determined by%
\begin{equation}
\tilde{\Omega}_{n,0}=\Omega_{n,0}-\omega_{s}^{\left(  n\right)  },
\end{equation}
where the Huang-Rhys factor $\omega_{s}^{\left(  n\right)  }$ describes the
energy shift due to lattice relaxation. The physical meaning of the parameters
$\omega_{s}^{\left(  n\right)  }$ obviously implies that the peaks at
$\omega=\tilde{\Omega}_{n,0}+k$ should be attributed to transitions to the RES
with emission of $k$ phonons. So, the so-called \textquotedblleft FC
transition\textquotedblright \ is realized as the envelope of a series of
phonon sidebands of the polaron RES but not as a transition to the FC state.
The account of the Jahn-Teller effects in general makes the multiphonon peak
series non-equidistant, but it changes nothing in the concept of the internal
polaron states which is discussed above.

\section{Conclusions \label{Conclusions}}

We have derived the polaron optical conductivity which is asymptotically exact
in the strong-coupling limit. The strong-coupling polaron optical conductivity
band is provided by the multiphonon transitions from the polaron ground state
to the polaron RES and has the maximum positioned close to the FC transition
frequency. With increasing the electron-phonon coupling constant $\alpha$, the
polaron optical conductivity band shape gradually tends to that provided by
the Diagrammatic Quantum Monte Carlo (DQMC) method. This agreement
demonstrates the importance of the multiphonon processes for the polaron
optical conductivity in the strong-coupling regime.

The obtained polaron optical conductivity with a high accuracy satisfies the
sum rule \cite{DLR1977}, what gives us an evidence of the fact that in the
strong-coupling regime the dominating contribution to the polaron optical
conductivity is due to the transitions to the \emph{internal} polaron states,
while the contribution due to the transitions to the continuum states is
negligibly small.

Accurate numerical results, obtained using DQMC method \cite{Mishchenko2003},
-- modulo the linewidths for sufficiently large $\alpha$ -- and the
analytically exact in the strong-coupling limit polaron optical conductivity
of the present work, as well as the analytical approximation of Ref.
\cite{DeFilippis2006} confirm the essence of the mechanism for the optical
absorption of Fr\"{o}hlich polarons, which were proposed in Refs.
\cite{DSG1972,Devreese1972}.

\begin{acknowledgments}
One of the authors (J.~T.~D.) thanks R. Evrard and E. Kartheuser for early
collaboration that led to Ref. \cite{KED1969} that contains key- conceptual
elements for the present work. J.~T.~D also thanks A.~S. Mishchenko for
providing his (unpublished) DQMC results for large coupling and for
stimulating discussions related to his DQMC. We thank F. Brosens, H. Kleinert,
G. Iadonisi, V. M. Fomin, G. De Filippis and V. Cataudella for valuable
discussions. This work was supported by FWO-V projects G.0356.06, G.0370.09N,
G.0180.09N, G.0365.08, the WOG WO.035.04N (Belgium).
\end{acknowledgments}

\appendix

\section{Correlation function}

The dipole-dipole correlation function $f_{zz}\left(  t\right)  $ given by
(\ref{fzz4}) is further simplified within the adiabatic approximation and
using the selection rules for the dipole transition matrix elements and the
symmetry properties of the polaron Hamiltonian. First, according to the
selection rules, the matrix element\textrm{ }$\left \langle \psi_{0}\left \vert
z\right \vert \psi_{n,l,m}\right \rangle \ $is
\begin{equation}
\left \langle \psi_{n^{\prime},l^{\prime},m^{\prime}}\left \vert z\right \vert
\psi_{0}\right \rangle =\delta_{l^{\prime},1}\delta_{m^{\prime},0}\left \langle
\psi_{n^{\prime},1,0}\left \vert z\right \vert \psi_{0}\right \rangle \label{b1}%
\end{equation}
\textrm{ }

Second, the interaction Hamiltonian $W$ (and hence, also the evolution
operator which involves $W$) is a scalar of the rotation symmetry group. The
matrix elements $\left \langle \psi_{n,l,m}\left \vert W\left(  s\right)
\right \vert \psi_{n,l^{\prime},m^{\prime}}\right \rangle $ for $l\neq
l^{\prime}$ and $m\neq m^{\prime}$ are then exactly equal to zero. Therefore,
in the adiabatic approximation and due to the symmetry of the Hamiltonian
$\tilde{H}$, we obtain the relations%
\begin{align}
&  \left \langle \psi_{0}\left \vert \mathrm{T}\exp \left(  i\int_{0}%
^{t}dsW\left(  -s\right)  \right)  \right \vert \psi_{n,l,m}\right \rangle
\nonumber \\
&  \approx \delta_{n,0}\delta_{l,0}\delta_{m,0}\left \langle \psi_{0}\left \vert
\mathrm{T}\exp \left(  -i\int_{0}^{t}dsW\left(  s\right)  \right)  \right \vert
\psi_{0}\right \rangle , \label{b2}%
\end{align}%
\begin{align}
&  \left \langle \psi_{n^{\prime \prime},l^{\prime \prime},m^{\prime \prime}%
}\left \vert \mathrm{T}\exp \left(  -i\int_{0}^{t}dsW\left(  s\right)  \right)
\right \vert \psi_{n^{\prime},l^{\prime},m^{\prime}}\right \rangle \nonumber \\
&  \approx \delta_{n^{\prime \prime},n^{\prime}}\delta_{l^{\prime \prime
},l^{\prime}}\left \langle \psi_{n^{\prime},l^{\prime},m^{\prime}}\left \vert
\mathrm{T}\exp \left(  -i\int_{0}^{t}dsW\left(  s\right)  \right)  \right \vert
\psi_{n^{\prime},l^{\prime},m^{\prime}}\right \rangle . \label{b3}%
\end{align}
Furthermore, because the ground state $\psi_{0}$ is non-degenerate, we find
that
\[
\left \langle \psi_{0}\left \vert \mathrm{T}\exp \left(  -i\int_{0}^{t}dsW\left(
s\right)  \right)  \right \vert \psi_{0}\right \rangle \approx1,
\]
because within the adiabatic approximation, for any $n\geq1$ the averages
$\left \langle \psi_{0}\left \vert W^{n}\right \vert \psi_{0}\right \rangle =0$.

The correlation function (\ref{fzz4}) using (\ref{b1}) to (\ref{b3}) takes the
form%
\begin{align}
f_{zz}\left(  t\right)   &  =\sum_{n}D_{n}e^{-i\Omega_{n,0}t}\nonumber \\
&  \times \left \langle \psi_{n,1,0}\left \vert \left \langle 0_{ph}\left \vert
\mathrm{T}\exp \left[  -i\int_{0}^{t}dsW\left(  s\right)  \right]  \right \vert
0_{ph}\right \rangle \right \vert \psi_{n,1,0}\right \rangle \label{f-ad}%
\end{align}
with the squared matrix elements of the dipole transitions%
\begin{equation}
D_{n}\equiv \left \vert \left \langle \psi_{n,1,0}\left \vert z\right \vert
\psi_{0}\right \rangle \right \vert ^{2}=\frac{1}{3}\left(  \int_{0}^{\infty
}R_{n,1}\left(  r\right)  R_{0,0}\left(  r\right)  r^{3}dr\right)  ^{2},
\label{dd}%
\end{equation}
and the FC transition frequencies%
\begin{equation}
\Omega_{n,0}\equiv E_{n,1}-E_{0}. \label{FC}%
\end{equation}
Further on, the interaction Hamiltonian is expressed in terms of the complex
phonon coordinates $Q_{\mathbf{k}}$:%
\begin{equation}
W=\sqrt{2}\sum_{\mathbf{k}}W_{\mathbf{k}}Q_{\mathbf{k}},\quad Q_{\mathbf{k}%
}=\frac{b_{\mathbf{k}}+b_{-\mathbf{k}}^{+}}{\sqrt{2}} \label{W3}%
\end{equation}
Here, we use the spherical-wave basis for phonon modes:%
\begin{equation}
\varphi_{k,l,m}\left(  \mathbf{r}\right)  \equiv \left(  -1\right)
^{\frac{m-\left \vert m\right \vert }{2}}\phi_{k,l}\left(  r\right)
Y_{l,m}\left(  \theta,\varphi \right)  ,
\end{equation}
where the radial part of the basis function is expressed through the spherical
Bessel function $j_{l}\left(  kr\right)  $:
\begin{equation}
\phi_{k,l}\left(  r\right)  =\left(  \frac{2}{R}\right)  ^{1/2}k\ j_{l}\left(
kr\right)  ,\;R=\left(  \frac{3V}{4\pi}\right)  ^{1/3}.
\end{equation}
The factor $\left(  -1\right)  ^{\frac{m-\left \vert m\right \vert }{2}}$ is
chosen in order to fulfil the symmetry property%
\[
\varphi_{k,l,m}^{\ast}\left(  \mathbf{r}\right)  =\varphi_{k,l,-m}\left(
\mathbf{r}\right)  .
\]
In the spherical-wave basis, the interaction Hamiltonian is%
\begin{equation}
W=\sqrt{2}\sum_{k,l,m}W_{k,l,m}Q_{k,l,m}, \label{W4}%
\end{equation}
with the complex phonon coordinates%
\begin{equation}
Q_{k,l,m}=\frac{b_{k,l,m}+b_{k,l,-m}^{+}}{\sqrt{2}} \label{Q}%
\end{equation}
and with the interaction amplitudes%
\begin{equation}
W_{k,l,m}=\frac{\sqrt{2\sqrt{2}\pi \alpha}}{k}\left(  \varphi_{k,l,m}\left(
\mathbf{r}\right)  -\rho_{k,l,m}\right)  ,\; \rho_{k,l,m}\equiv \left \langle
\psi_{0}\left \vert \varphi_{k,l,m}\right \vert \psi_{0}\right \rangle .
\end{equation}
The dipole-dipole correlation function (\ref{f-ad}) is then%
\begin{align}
f_{zz}\left(  t\right)   &  =\sum_{n}D_{n}e^{-i\Omega_{n,0}t}\nonumber \\
&  \times \left \langle \psi_{n,1,0}\left \vert \left \langle 0_{ph}\left \vert
\mathrm{T}\exp \left[  -i\sqrt{2}\int_{0}^{t}ds\sum_{k,l,m}W_{k,l,m}\left(
s\right)  Q_{k,l,m}\left(  s\right)  \right]  \right \vert 0_{ph}\right \rangle
\right \vert \psi_{n,1,0}\right \rangle . \label{f-ad1}%
\end{align}
The operators $W_{k,l,m}\left(  s\right)  $ in (\ref{f-ad1}) are equivalent to
the $\left(  2l+1\right)  $-dimensional matrices $\mathbb{W}_{k,l,m}^{\left(
n\right)  }$ determined in the basis of the level $\left(  n,l\right)  $. The
matrix elements of these matrices are%
\begin{equation}
\left(  \mathbb{W}_{k,l,m}^{\left(  n\right)  }\right)  _{m_{1},m_{2}%
}=\left \langle \psi_{n,1,m_{1}}\left \vert W_{k,l,m}\right \vert \psi
_{n,1,m_{2}}\right \rangle . \label{matr}%
\end{equation}
In these notations, $f_{zz}\left(  t\right)  $ given by (\ref{f-ad1}) can be
written down as%
\begin{equation}
f_{zz}\left(  t\right)  =\sum_{n}D_{n}e^{-i\Omega_{n,0}t}\left \langle
0_{ph}\left \vert \left(  \mathrm{T}\exp \left[  -i\int_{0}^{t}ds\mathbb{W}%
^{\left(  n\right)  }\left(  s\right)  \right]  \right)  _{0,0}\right \vert
0_{ph}\right \rangle . \label{f-ad2}%
\end{equation}
where $\mathbb{W}^{\left(  n\right)  }$ is the matrix electron-phonon
interaction Hamiltonian expressed through the phonon complex coordinates in
the spherical-wave representation as follows:%
\begin{equation}
\mathbb{W}^{\left(  n\right)  }=\sqrt{2}\sum_{k,l,m}\mathbb{W}_{k,l,m}%
^{\left(  n\right)  }Q_{k,l,m}. \label{WM}%
\end{equation}
Here, $\mathbb{W}_{k,l,m}^{\left(  n\right)  }$ is a $\left(  3\times3\right)
$ matrix in a basis of a level $\left(  n,l\right)  _{l=1}$ of the Hamiltonian
$\tilde{H}_{0}$.

Because $\mathbb{W}^{\left(  n\right)  }$ is a scalar of the rotation group,
we can replace the diagonal matrix element of the T-exponent in (\ref{f-ad2})
with the trace in the aforesaid-finite-dimensional basis. As a result, we
obtain for the polaron optical conductivity (\ref{KuboDD0}) with (\ref{f-ad2})
the expression%
\begin{align}
\operatorname{Re}\sigma \left(  \omega \right)   &  =\frac{\omega}{6}\sum
_{n}D_{n}\int_{-\infty}^{\infty}e^{i\left(  \omega-\Omega_{n,0}\right)
t}\nonumber \\
&  \times \left \langle 0_{ph}\left \vert \mathrm{Tr}\left(  \mathrm{T}%
\exp \left[  -i\int_{0}^{t}ds\mathbb{W}^{\left(  n\right)  }\left(  s\right)
\right]  \right)  \right \vert 0_{ph}\right \rangle dt. \label{f-ad3a}%
\end{align}

\section{Effective phonon modes}

In order to perform the averaging in Eq. (\ref{fzz1}) analytically, we
introduce the effective phonon modes $Q_{0,0}$ and $Q_{2,m}$ similarly to Ref.
\cite{Lumin1998}. The Hamiltonian $\mathbb{W}^{\left(  n\right)  }$ in terms
of these effective phonon modes is expressed as%
\begin{equation}
\mathbb{W}^{\left(  n\right)  }=\sqrt{2}\sum_{l,m}\mathbb{\tilde{W}}%
_{l,m}^{\left(  n\right)  }Q_{l,m} \label{W2a}%
\end{equation}
where the matrices $\mathbb{\tilde{W}}_{l,m}^{\left(  n\right)  }$ (depending
on the vibration coordinates $Q_{l,m}$) are explicitly given by the
expressions (cf. Ref. \cite{Lumin1998}),%
\begin{equation}
\mathbb{W}^{\left(  n\right)  }=a_{0}^{\left(  n\right)  }\mathbb{I}%
Q_{0,0}+a_{2}^{\left(  n\right)  }\sum_{m=-2}^{2}\mathbb{B}_{m}Q_{2,m}
\label{W2}%
\end{equation}
with the matrices $\mathbb{B}_{j}$
\begin{equation}
\mathbb{B}_{0}=\frac{1}{2\sqrt{5\pi}}\left(
\begin{array}
[c]{ccc}%
-1 & 0 & 0\\
0 & 2 & 0\\
0 & 0 & -1
\end{array}
\right)  , \label{B0}%
\end{equation}%
\begin{equation}
\mathbb{B}_{1}=\mathbb{B}_{-1}^{+}=\frac{1}{2}\sqrt{\frac{3}{5\pi}}\left(
\begin{array}
[c]{ccc}%
0 & 0 & 0\\
-1 & 0 & 0\\
0 & 1 & 0
\end{array}
\right)  , \label{B1}%
\end{equation}%
\begin{equation}
\mathbb{B}_{2}=\mathbb{B}_{-2}^{+}=\sqrt{\frac{3}{10\pi}}\left(
\begin{array}
[c]{ccc}%
0 & 0 & 0\\
0 & 0 & 0\\
-1 & 0 & 0
\end{array}
\right)  . \label{B2}%
\end{equation}
The coefficients $a_{0}^{\left(  n\right)  }$ and $a_{2}^{\left(  n\right)  }$
in Eq. (\ref{W2}) are%
\begin{align}
a_{0}^{\left(  n\right)  }  &  =\left(  \sqrt{2}\alpha \sum_{k}\frac{1}{k^{2}%
}\left[  \left \langle \phi_{k,0}\right \rangle _{n,1}-\left \langle \phi
_{k,0}\right \rangle _{0,0}\right]  ^{2}\right)  ^{1/2},\label{c0}\\
a_{2}^{\left(  n\right)  }  &  =\left(  4\sqrt{2}\pi \alpha \sum_{k}\frac
{1}{k^{2}}\left \langle \phi_{k,2}\right \rangle _{n,1}^{2}\right)  ^{1/2}.
\label{c2}%
\end{align}
Here $\phi_{k,l}$ is the radial part of the basis function expressed through
the spherical Bessel function $j_{l}\left(  kr\right)  $:
\begin{equation}
\phi_{k,l}\left(  r\right)  =\left(  \frac{2}{R}\right)  ^{1/2}k\ j_{l}\left(
kr\right)  ,\;R=\left(  \frac{3V}{4\pi}\right)  ^{1/3}, \label{Rad}%
\end{equation}
$V$ is the volume of the crystal, and $\left \langle f\left(  r\right)
\right \rangle _{n,l}$ is the average%
\begin{equation}
\left \langle f\left(  r\right)  \right \rangle _{n,l}=\int_{0}^{\infty}f\left(
r\right)  R_{n,l}^{2}\left(  r\right)  r^{2}dr.
\end{equation}
The normalization of the phonon wave functions corresponds to the condition%
\begin{equation}
\int_{0}^{R}\phi_{k,l}\left(  r\right)  \phi_{k^{\prime},l}\left(  r\right)
r^{2}dr=\delta_{k,k^{\prime}}. \label{norm}%
\end{equation}
After the straightforward calculation using (\ref{norm}), we express the
coefficients $a_{0}^{\left(  n\right)  }$ and $a_{2}^{\left(  n\right)  }$
through the integrals with the radial wave functions:%
\begin{align}
a_{0}^{\left(  n\right)  }  &  =\left(  2\sqrt{2}\alpha \int_{0}^{\infty}%
dr\int_{0}^{r}dr^{\prime}\ r\left(  r^{\prime}\right)  ^{2}\left[  R_{n,1}%
^{2}\left(  r\right)  -R_{0,0}^{2}\left(  r\right)  \right]  \left[
R_{n,1}^{2}\left(  r^{\prime}\right)  -R_{0,0}^{2}\left(  r^{\prime}\right)
\right]  \right)  ^{1/2},\label{c0n}\\
a_{2}^{\left(  n\right)  }  &  =\left(  \frac{8\sqrt{2}\pi \alpha}{5}\int
_{0}^{\infty}dr\int_{0}^{r}dr^{\prime}\frac{\left(  r^{\prime}\right)  ^{4}%
}{r}R_{n,1}^{2}\left(  r\right)  R_{n,1}^{2}\left(  r^{\prime}\right)
\right)  ^{1/2}. \label{c2n}%
\end{align}

\subsection{Exact averaging}

Let us substitute the matrix interaction Hamiltonian (\ref{W2}) to the
dipole-dipole correlation function (\ref{fzz1}), what gives us the result%
\begin{equation}
f_{zz}\left(  t\right)  =\frac{1}{3}\sum_{n}D_{n}e^{-i\Omega_{n,0}%
t}\left \langle 0_{ph}\left \vert \exp \left(  -ita_{0}^{\left(  n\right)  }%
Q_{0}\right)  \mathrm{Tr}\exp \left(  -it\frac{a_{2}^{\left(  n\right)  }%
}{2\sqrt{5\pi}}\mathbb{V}\left(  Q_{2}\right)  \right)  \right \vert
0_{ph}\right \rangle . \label{fzz2a}%
\end{equation}
Here, we use the matrix depending on the phonon coordinates,%
\begin{equation}
\mathbb{V}\left(  Q_{2}\right)  \equiv2\sqrt{5\pi}\sum_{m=-2}^{2}%
\mathbb{B}_{m}Q_{2m}, \label{VQ}%
\end{equation}
whose explicit form is%
\begin{equation}
\mathbb{V}\left(  Q_{2}\right)  =\left(
\begin{array}
[c]{ccc}%
-Q_{2,0} & -\sqrt{3}Q_{2,-1} & -\sqrt{6}Q_{2,-2}\\
-\sqrt{3}Q_{2,1} & 2Q_{2,0} & \sqrt{3}Q_{2,-1}\\
-\sqrt{6}Q_{2,2} & \sqrt{3}Q_{2,1} & -Q_{2,0}%
\end{array}
\right)  . \label{MatrA}%
\end{equation}

The matrix $\mathbb{V}\left(  Q_{2}\right)  $ is analytically diagonalized.
The equation for the eigenvectors $\left \vert \chi \left(  Q_{2}\right)
\right \rangle $ and eigenvalues $\lambda \left(  Q_{2}\right)  $ of
$\mathbb{V}\left(  Q_{2}\right)  $ is%
\begin{equation}
\mathbb{V}\left(  Q_{2}\right)  \left \vert \chi \left(  Q_{2}\right)
\right \rangle =\lambda \left(  Q_{2}\right)  \left \vert \chi \left(
Q_{2}\right)  \right \rangle . \label{SL}%
\end{equation}
The eigenvalues are found from the equation%
\begin{equation}
\det \left(  \mathbb{V}\left(  Q_{2}\right)  -\lambda \left(  Q_{2}\right)
\mathbb{I}\right)  =0. \label{SE}%
\end{equation}
We make the transformation to the real phonon coordinates,
\begin{align*}
Q_{2,0}  &  \equiv x_{0},\\
Q_{2,m}  &  \equiv \frac{x_{m}+iy_{m}}{\sqrt{2}},\quad Q_{2,-m}=Q_{2,m}^{\ast
}=\frac{x_{m}-iy_{m}}{\sqrt{2}}.
\end{align*}
Five variables $x_{0},x_{1},x_{2},y_{1},y_{2}$ are the independent real phonon
coordinates. The l.h.s. of Eq. (\ref{SE}) is expressed in terms of these
coordinates as
\begin{equation}
\det \left(  \mathbb{V}\left(  Q_{2}\right)  -\lambda \left(  Q_{2}\right)
\mathbb{I}\right)  =-\lambda^{3}+3p\lambda+2q
\end{equation}
with the coefficients
\begin{align*}
p  &  =x_{0}^{2}+x_{1}^{2}+x_{2}^{2}+y_{1}^{2}+y_{2}^{2},\\
q  &  =x_{0}^{3}+\frac{3}{2}x_{0}\left(  x_{1}^{2}+y_{1}^{2}\right)
+\frac{3\sqrt{3}}{2}x_{2}\left(  x_{1}^{2}-y_{1}^{2}\right)  -3x_{0}\left(
x_{2}^{2}+y_{2}^{2}\right)  +3\sqrt{3}x_{1}y_{1}y_{2}.
\end{align*}
So, we have the cubic equation for $\lambda$:%
\begin{equation}
\lambda^{3}-3p\lambda-2q=0. \label{cub}%
\end{equation}
Because the matrix $\mathbb{V}\left(  Q_{2}\right)  $ is Hermitian, all its
eigenvalues are real. Therefore, $\frac{\left \vert q\right \vert }{p^{3/2}}%
\leq1$ (otherwise, $\sin \left(  3\varphi \right)  $ is not real). Herefrom, we
have three explicit eigenvalues:%
\begin{align}
\lambda_{1}\left(  Q_{2}\right)   &  =2\sqrt{p}\sin \left[  \frac{\pi}{3}%
+\frac{1}{3}\arcsin \left(  \frac{q}{p^{3/2}}\right)  \right]  ,\nonumber \\
\lambda_{2}\left(  Q_{2}\right)   &  =-2\sqrt{p}\sin \left[  \frac{1}{3}%
\arcsin \left(  \frac{q}{p^{3/2}}\right)  \right]  ,\nonumber \\
\lambda_{3}\left(  Q_{2}\right)   &  =-2\sqrt{p}\sin \left[  \frac{\pi}%
{3}-\frac{1}{3}\arcsin \left(  \frac{q}{p^{3/2}}\right)  \right]  .
\label{eigenv}%
\end{align}

The trace in (\ref{fzz2a}) is invariant with respect to the choice of the
basis. Consequently, after the diagonalization $f_{zz}\left(  t\right)  $
takes the form%
\begin{equation}
f_{zz}\left(  t\right)  =\frac{1}{3}\sum_{n}D_{n}e^{-i\Omega_{n,0}t}\sum
_{j=1}^{3}\left \langle 0_{ph}\left \vert \exp \left(  -it\left[  a_{0}^{\left(
n\right)  }Q_{0}+\frac{a_{2}^{\left(  n\right)  }}{2\sqrt{5\pi}}\lambda
_{j}\left(  Q_{2}\right)  \right]  \right)  \right \vert 0_{ph}\right \rangle .
\label{fzz3b}%
\end{equation}
After inserting $f_{zz}\left(  t\right)  $ given by (\ref{fzz3b}) into
(\ref{KuboDD0}), the integration over time gives the delta function multiplied
by $2\pi$, and we arrive at the result%
\begin{equation}
\operatorname{Re}\sigma \left(  \omega \right)  =\frac{\pi \omega}{3}\sum
_{n}D_{n}\sum_{j=1}^{3}\left \langle 0_{ph}\left \vert \delta \left(
\omega-\Omega_{n,0}-a_{0}^{\left(  n\right)  }Q_{0}-\frac{a_{2}^{\left(
n\right)  }}{2\sqrt{5\pi}}\lambda_{j}\left(  Q_{2}\right)  \right)
\right \vert 0_{ph}\right \rangle . \label{fzz5}%
\end{equation}
The ground-state wave function for the effective phonon modes is%
\begin{equation}
\left \vert 0_{ph}\right \rangle \equiv \Phi_{0}\left(  Q\right)  =\Phi
_{0}^{\left(  0\right)  }\left(  Q_{0}\right)  \Phi_{0}^{\left(  2\right)
}\left(  Q_{2}\right)  . \label{0ph}%
\end{equation}
$\Phi_{0}^{\left(  0\right)  }\left(  Q_{0}\right)  $ is the one-oscillator
ground-state wave function:
\begin{equation}
\Phi_{0}^{\left(  0\right)  }\left(  Q_{0}\right)  =\pi^{-1/4}\exp \left(
-\frac{Q_{0}^{2}}{2}\right)  .
\end{equation}
The ground-state wave function of phonons with $l=2$ is:
\begin{equation}
\Phi_{0}^{\left(  2\right)  }\left(  Q_{2}\right)  =\pi^{-5/4}\exp \left[
-\frac{1}{2}\left(  x_{0}^{2}+\sum_{m=1,2}\left(  x_{m}^{2}+y_{m}^{2}\right)
\right)  \right]  .
\end{equation}
The phonon ground-state wave function (\ref{0ph}) is then%
\begin{equation}
\Phi_{0}\left(  Q\right)  =\frac{1}{\pi^{3/2}}\exp \left[  -\frac{1}{2}\left(
x_{0}^{2}+\sum_{m=1,2}\left(  x_{m}^{2}+y_{m}^{2}\right)  +Q_{0}^{2}\right)
\right]  .
\end{equation}
With these phonon wave functions, Eq. (\ref{fzz5}) results in the following
expression for the polaron optical conductivity
\begin{align}
\operatorname{Re}\sigma \left(  \omega \right)   &  =\frac{\omega}{3\pi^{2}}%
\sum_{n}\frac{D_{n}}{a_{0}^{\left(  n\right)  }}\int_{-\infty}^{\infty}%
dx_{0}\int_{-\infty}^{\infty}dx_{1}\int_{-\infty}^{\infty}dx_{2}\int_{-\infty
}^{\infty}dy_{1}\int_{-\infty}^{\infty}dy_{2}\nonumber \\
&  \times \sum_{j=1}^{3}\exp \left \{  -\frac{1}{2}\left[  x_{0}^{2}+\sum
_{m=1,2}\left(  x_{m}^{2}+y_{m}^{2}\right)  +\frac{\left(  \omega-\Omega
_{n,0}-\frac{a_{2}^{\left(  n\right)  }}{2\sqrt{5\pi}}\lambda_{j}\left(
Q_{2}\right)  \right)  ^{2}}{\left(  a_{0}^{\left(  n\right)  }\right)  ^{2}%
}\right]  \right \}  . \label{Resig1}%
\end{align}

\subsection{Averaging neglecting the Jahn-Teller effect}

In order to perform the phonon averaging explicitly, we disentangle the
exponent $\exp \left(  -it\sqrt{2}\sum_{l,m}\mathbb{\tilde{W}}_{l,m}^{\left(
n\right)  }Q_{l,m}\right)  $ as follows.%
\begin{align}
&  \exp \left(  -it\sqrt{2}\sum_{l,m}\mathbb{\tilde{W}}_{l,m}^{\left(
n\right)  }Q_{l,m}\right)  =\exp \left(  -it\sum_{l,m}\mathbb{\tilde{W}}%
_{l,-m}^{\left(  n\right)  }b_{l,m}^{+}\right) \nonumber \\
&  \times \mathrm{T}\exp \left(  -i\int_{0}^{t}ds\sum_{l,m}e^{is\sum_{l^{\prime
},m^{\prime}}\mathbb{\tilde{W}}_{l^{\prime},-m^{\prime}}^{\left(  n\right)
}b_{l^{\prime},m^{\prime}}^{+}}\mathbb{\tilde{W}}_{l,m}^{\left(  n\right)
}b_{l,m}e^{-is\sum_{l^{\prime},m^{\prime}}\mathbb{\tilde{W}}_{l^{\prime
},-m^{\prime}}^{\left(  n\right)  }b_{l^{\prime},m^{\prime}}^{+}}\right)  .
\end{align}

Neglecting non-commutation of matrices $\mathbb{\tilde{W}}_{l,m}^{\left(
n\right)  }$ we find that%
\begin{align}
&  \sum_{l,m}e^{is\sum_{l^{\prime},m^{\prime}}\mathbb{\tilde{W}}_{l^{\prime
},-m^{\prime}}^{\left(  n\right)  }b_{l^{\prime},m^{\prime}}^{+}%
}\mathbb{\tilde{W}}_{l,m}^{\left(  n\right)  }b_{l,m}e^{-is\sum_{l^{\prime
},m^{\prime}}\mathbb{\tilde{W}}_{l^{\prime},-m^{\prime}}^{\left(  n\right)
}b_{l^{\prime},m^{\prime}}^{+}}\nonumber \\
&  =\sum_{l,m}\mathbb{\tilde{W}}_{l,m}^{\left(  n\right)  }b_{l,m}%
-is\sum_{l,m}\mathbb{\tilde{W}}_{l,-m}^{\left(  n\right)  }\mathbb{\tilde{W}%
}_{l,m}^{\left(  n\right)  }.
\end{align}

The sum $\sum_{l,m}\mathbb{\tilde{W}}_{l,-m}^{\left(  n\right)  }%
\mathbb{\tilde{W}}_{l,m}^{\left(  n\right)  }$ in the basis ($l,m$) for a
definite $n$ is proportional to the unity matrix. Therefore, $\exp \left(
-it\sqrt{2}\sum_{l,m}\mathbb{\tilde{W}}_{l,m}^{\left(  n\right)  }%
Q_{l,m}\right)  $ is%
\begin{align}
&  e^{-it\sqrt{2}\sum_{l,m}\mathbb{\tilde{W}}_{l,m}^{\left(  n\right)
}Q_{l,m}}\nonumber \\
&  =e^{-it\sum_{l,m}\mathbb{\tilde{W}}_{l,-m}^{\left(  n\right)  }b_{l,m}^{+}%
}e^{-it\sum_{l,m}\mathbb{\tilde{W}}_{l,m}^{\left(  n\right)  }b_{l,m}%
-\frac{t^{2}}{2}\sum_{l,m}\mathbb{\tilde{W}}_{l,-m}^{\left(  n\right)
}\mathbb{\tilde{W}}_{l,m}^{\left(  n\right)  }},
\end{align}
that gives us the result%
\begin{equation}
\left \langle 0_{ph}\left \vert e^{-it\sqrt{2}\sum_{l,m}\mathbb{\tilde{W}}%
_{l,m}^{\left(  n\right)  }Q_{l,m}}\right \vert 0_{ph}\right \rangle
=e^{-\frac{t^{2}}{2}\sum_{l,m}\mathbb{\tilde{W}}_{l,-m}^{\left(  n\right)
}\mathbb{\tilde{W}}_{l,m}^{\left(  n\right)  }}.
\end{equation}
Using the explicit formulae for the matrices $\tilde{W}_{l,m}^{\left(
n\right)  }$, the matrix sum takes the form%
\begin{equation}
\sum_{l,m}\mathbb{\tilde{W}}_{l,-m}^{\left(  n\right)  }\mathbb{\tilde{W}%
}_{l,m}^{\left(  n\right)  }=\omega_{s}^{\left(  n\right)  }\mathbb{I}
\label{exp1}%
\end{equation}
with the parameter%
\begin{equation}
\omega_{s}^{\left(  n\right)  }=\frac{1}{2}\left(  a_{0}^{\left(  n\right)
}\right)  ^{2}+\frac{1}{4\pi}\left(  a_{2}^{\left(  n\right)  }\right)  ^{2}.
\end{equation}
Using (\ref{exp1}), the optical conductivity (\ref{fzz1}) is transformed to
the expression%
\begin{equation}
\operatorname{Re}\sigma \left(  \omega \right)  =\omega \sum_{n}\sqrt{\frac{\pi
}{2S_{n}}}D_{n}\exp \left(  -\frac{\left(  \omega-\Omega_{n,0}\right)  ^{2}%
}{2S_{n}}\right)  . \label{ReS5}%
\end{equation}

\newpage

\textbf{Figure captions}

\bigskip

Fig. 1. Frequency of the main peak in the optical conductivity spectra
calculated within the model of Ref. \cite{DSG1972} (dots) and the main-peak
energy extracted from the DQMC data \cite{Mishchenko2003,DeFilippis2006} (squares).

\bigskip

Fig. 2. The strong-coupling polaron optical conductivity calculated within the
rigorous strong-coupling approach of the present work (solid curves), within
the present approach but neglecting the Jahn -- Teller effect (dashed curves),
within the adiabatic approximation of Ref. \cite{DeFilippis2006} (dot-dashed
curves), and the numerical Diagrammatic Monte Carlo data (full dots) for
$\alpha=8$ and 8.5. The FC and RES transition frequencies are indicated by the arrows.

\bigskip

Fig. 3. The strong-coupling polaron optical conductivity calculated within the
rigorous strong-coupling approach of the present work (solid curves), within
the present approach but neglecting the Jahn -- Teller effect (dashed curves),
within the adiabatic approximation of Ref. \cite{DeFilippis2006} (dot-dashed
curves), and the numerical Diagrammatic Monte Carlo data (full dots) for
$\alpha=9$, 13 and 15. The FC and RES transition frequencies are indicated by
the arrows.

\newpage

\textbf{Figures}

\bigskip%

\begin{figure}
[h]
\begin{center}
\includegraphics[
height=4.4324in,
width=4.3457in
]%
{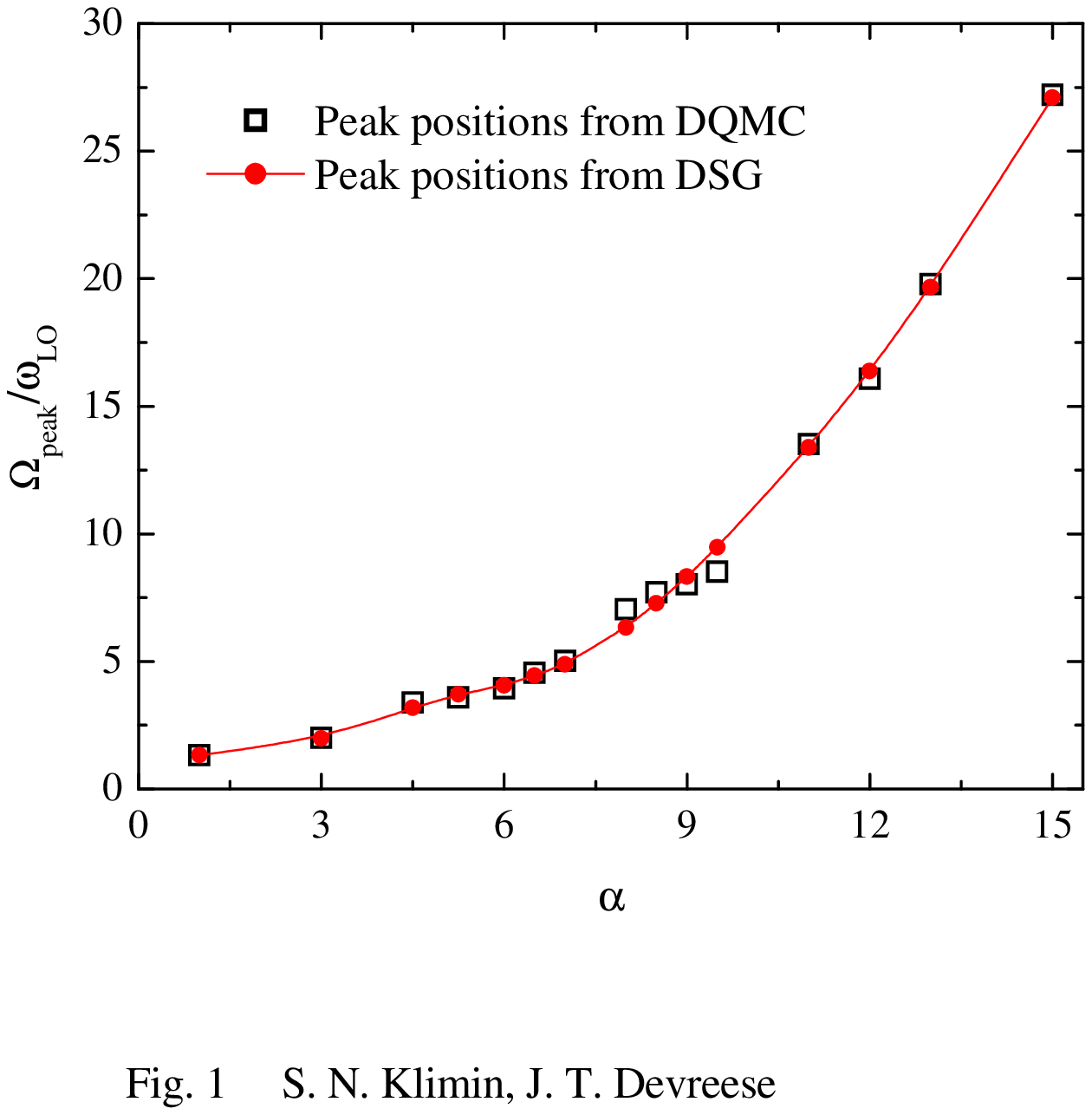}%
\end{center}
\end{figure}

\newpage%

\begin{figure}
[h]
\begin{center}
\includegraphics[
height=5.2239in,
width=3.3181in
]%
{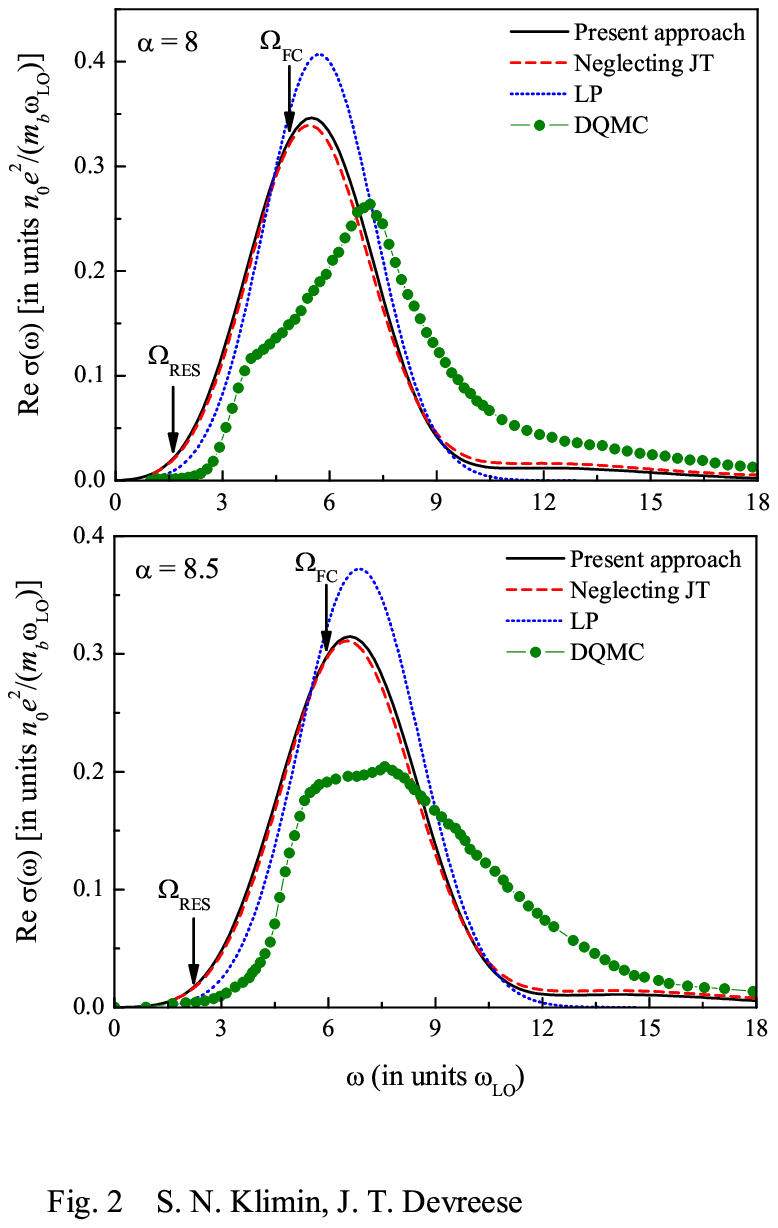}%
\end{center}
\end{figure}

\newpage%

\begin{figure}
[h]
\begin{center}
\includegraphics[
height=7.3041in,
width=3.3412in
]%
{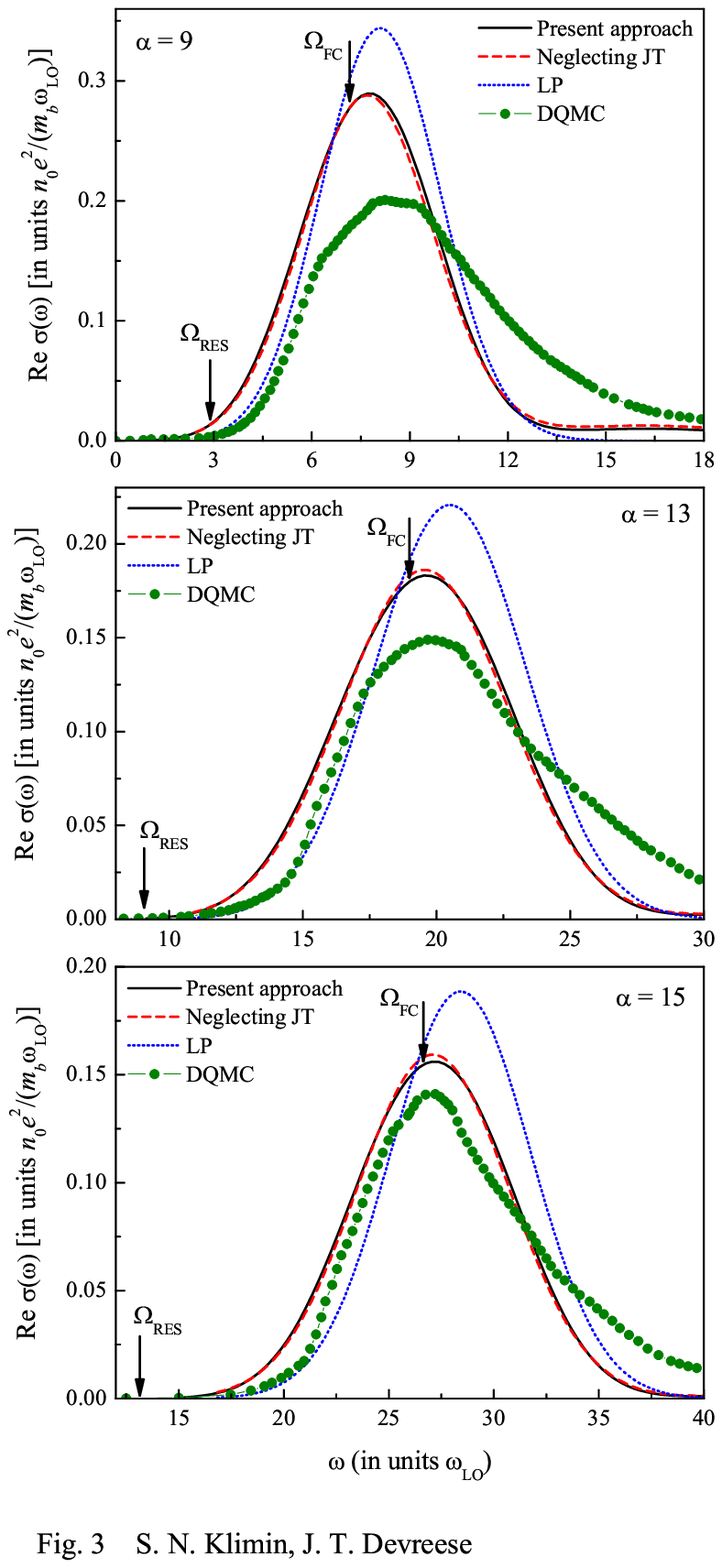}%
\end{center}
\end{figure}

\end{document}